\documentclass[superbib,aps,twocolumn]{revtex4-1}
\usepackage[english]{babel}
\usepackage{amsmath,amsfonts,amssymb}
\usepackage[pdftex]{graphicx}
\usepackage{bm} % bold math
\begin{document}
%The title:
\title{Dense electron-positron plasmas and bursts of gamma-rays from laser-generated QED plasmas}
%Author(s)
\author{C.P. Ridgers$^1$, C.S.Brady$^{2}$, R. Duclous$^3$, J.G. Kirk$^4$, K. Bennett$^2$, T.D. Arber$^2$, A.R. Bell$^1$}

\address{{\small{\small{$^1$Clarendon Laboratory, University of Oxford, Parks Road, Oxford, OX1 3PU, UK \\
$^2$Centre for Fusion, Space and Astrophysics, University of Warwick, Coventry, CV4 7AL, UK \\
$^3$Commissariat \`{a} l'Energie Atomique, DAM DIF, F-91297 Arpajon, France  \\
$^4$Max-Planck-Institut f\"{u}r Kernphysik, Postfach 10 39 80, 69029 Heidelberg, Germany}}}}

\date{\today}

\begin{abstract}
In simulations of a 12.5PW laser (focussed intensity $I=4\times10^{23}$Wcm$^{-2}$) striking a solid aluminum target $10\%$ of the laser energy is converted to gamma-rays.  A dense electron-positron plasma is generated with a maximum density of $10^{26}$m$^{-3}$; seven orders of magnitude denser than pure e$^-$e$^+$ plasmas generated with 1PW lasers.  When the laser power is increased to 320PW ($I=10^{25}$Wcm$^{-2}$) 40\% of the laser energy is converted to gamma-ray photons and 10\% to electron-positron pairs.  In both cases there is strong feedback between the QED emission processes and the plasma physics; the defining feature of the new `QED-plasma' regime reached in these interactions.          
\end{abstract}

\maketitle

\section{Introduction}

%DEFINE ALL VARIABLES
%CHECK CONSISTENCY WITH PRL
%SEPARATE ZHIDKOV

With the advent of next generation 10PW-100PW lasers \cite{Mourou_07} a new frontier will be reached in high-power laser-plasma physics. These lasers will create strong enough electromagnetic fields to access strong-field quantum electrodynamics (QED) processes thought to be responsible for cascades of antimatter production in the relativistic winds from pulsars and black holes \cite{Goldreich_69}.  Strong-field QED processes have typically been investigated using particle accelerators in experiments arranged such that these QED scattering processes can be studied in isolation \cite{Bula_96} (and their cross-sections compared to QED calculations \cite{Erber_66}). This is also true of laser-solid experiments where photon and pair production occur in the electric fields of the nuclei of high-$Z$ materials far from the laser focus \cite{Chen_10}. By contrast, the fields in a $>10$PW laser's focus will cause strong-field QED reactions directly \cite{Bell_08}.  In this case the QED processes strongly modify the basic plasma dynamics.  Conversely, the rates of the QED reactions depend on the electromagnetic fields which are determined by the plasma dynamics.  As a result of this feedback neither the QED nor the plasma physics may be considered in isolation, but both must be treated self-consistently in the resulting `QED-plasma' \cite{Ridgers_12}. 

The important strong-field QED emission processes are \cite{Erber_66}: (1) quantum-corrected synchrotron radiation; (2) multiphoton Breit-Wheeler pair production. In (1) electrons and positrons in the plasma radiate energetic gamma-ray photons when accelerated by the electromagnetic fields of the laser. In process (2) these photons interact with the laser fields and generate electron-positron pairs.  The controlling parameter for these processes is $\eta\approx\gamma\theta\sqrt{I/I_s}$, where $I$ is the laser intensity \& $I_s=\epsilon_0{}cE_s^2/2=2\times10^{29}$Wcm$^{-2}$ is the laser intensity at which the average field in the laser focus is equal to the Schwinger field $E_s=1.3\times10^{18}$Vm$^{-1}$; i.e. the field required to break down the vacuum into electron positron-pairs \cite{Sauter_31}.  $\theta\in[0,2]$ depends on the interaction geometry.  As $\eta$ reaches 0.1 (for an optical laser) the electrons in the plasma radiate a significant fraction of their energy as gamma-ray photons; the plasma enters the `radiation dominated' regime \cite{Sokolov_10}.  In this regime the radiation reaction force \cite{Dirac_38} must be included in the equation of motion of the electrons \& positrons.  As laser intensities increase from the current maximum of $10^{22}$Wcm$^{-2}$ \cite{Bahk_04} to exceed $I\sim10^{23}$Wcm$^{-2}$ the ratio $I/I_s$ and the $\gamma$ factor to which the laser accelerates the electrons increase in step and $\eta=0.1$ is reached.  10PW lasers should be able to push well into this regime; if all the energy of a 10PW laser pulse is focussed into a laser spot of radius one micron $I=3\times10^{23}$Wcm$^{-2}$. When $\eta=1$ the following quantum corrections to the gamma-ray radiation become important: (1) the gamma-ray photon spectrum is modified to account for the recoil of the electron as it emits \cite{Erber_66}; (2) the emitted photon energy becomes a significant fraction of the emitting particles energy and therefore the emission becomes stochastic \cite{Shen_72}.  In addition a significant amount of the radiated photons generate electron-positron pairs.  Therefore when $\eta=1$ the `QED dominated' regime is reached \cite{Sokolov_10}. In optical laser-plasma interactions this will occur as intensities increase to $I\sim10^{24}$Wcm$^{-2}-10^{25}$Wcm$^{-2}$, the limit attainable with 30PW-300PW lasers.

Due to the complexity of the feedback between the QED emission processes and plasma physics effects in a realistic laser-produced QED-plasma, numerical simulations of these interactions are essential.  The appropriate simulation tool is obtained by augmenting a particle-in-cell (PIC) code by including QED emission processes (1) \& (2) above.  A classical model of gamma-ray emission and the resulting radiation reaction has previously been included in several PIC codes \cite{Zhidkov_02,Chen_11,Nakamura_12}.  A classical model is only valid in the relatively narrow intensity range defining the radiation dominated regime.  A quantum treatment of gamma-ray photon and pair generation, valid in the radiation and QED dominated regimes, has only recently been coupled to PIC codes \cite{Ridgers_12,Sokolov_09,Timhokin_10,Nerush_11}.  

The resulting QED-PIC codes have been used to self-consistently simulate cascades of electron-positron pair production, where a critical density pair plasma can be generated from a single seed electron, in pulsar atmospheres \cite{Timhokin_10} and in the interaction of two counter-propagating 100PW ($I=3\times10^{24}$Wcm$^{-2}$) laser pulses \cite{Nerush_11}.  QED-PIC simulations have also recently been used to show that prolific gamma-ray photon and pair production is possible in 10PW laser-solid interactions \cite{Ridgers_12}.  Here the interaction of the laser pulse with a dense plasma, combined with the fact that the solid reflects the laser pulse and so doubles the electric field, compensates for the expected QED rate reductions due to the ten times lower intensity.  In this paper we present a study of the interaction of $O$(10PW) and $O$(100PW) laser pulses with solid aluminum targets, exploring this promising configuration for pair production.  We will use QED-PIC {\small{EPOCH}} \cite{Brady_11} simulations to elucidate the details of the feedback between QED emission processes and plasma physics effects.  In doing so we will attempt to outline a theoretical framework for laser-solid interactions in the QED-plasma regime, which will be important not only for determining the most effective way to produce copious numbers of gamma-ray photons and pairs, but also in determining the viability of any proposed applications of $>$10PW laser-solid interactions such as ion acceleration to multi-GeV energies \cite{Esirkepov_04,Robinson_09} and high-harmonic generation \cite{Dromey_06}.

\section{QED Emission Model}

In this section we will discuss the model used for the QED emission processes its coupling to a PIC code.  This is dramatically simplified by the fact that in high intensity laser-plasma interactions the macroscopic laser fields are effectively unchanged in the QED interactions \cite{Heinzl_11}.  In this case the laser's electromagnetic field may be treated classically \cite{Bagrov_90} and the QED reactions in the strong-field QED framework \cite{Furry_51}.  Two approximations may be made concerning the classical `macroscopic' laser fields.  (1) They are quasi-static. The length scale over which photons or pairs are formed is a factor of $a=eE_l\lambda_l/2\pi m_e c^2$ ($E_l$ is the electric field of the laser) times smaller than the laser wavelength $\lambda_l$ \cite{Kirk_09}.  For $I>10^{23}$Wcm$^{-2}$, $a\gg 1$, and the laser's fields may be treated as constant during the QED emission processes.  (2) The laser's fields are much weaker than the Schwinger field.  In this case the QED reaction rates in the general fields in the plasma are the same as those in plane wave fields and depend only on the Lorentz-invariant parameters: $\eta=(e\hbar/m_e^3c^4)|F_{\mu\nu}p^{\nu}|$ and $\chi=(e\hbar^22m_e^3c^4)|F_{\mu\nu}k^{\nu}|$ \cite{Erber_66}.  $p^{\mu}$ ($k^{\mu}$) is the electron's (photon's) 4-momentum.  For $I<10^{25}$Wcm$^{-2}$, $E/E_s<5\times10^{-3}$ and so approximation (2) holds.

Physically $\eta$ is the field perpendicular to the electron motion boosted into its rest frame.  This can be seen clearly by writing $\eta$ in terms of three-vectors in the ultra-relativistic limit: $\eta\approx{}(\gamma/E_s)|\mathbf{E}_{\perp}+\bm{\beta}\times{}c\mathbf{B}|$; similarly $\chi=(\hbar\omega_{\gamma}/2m_ec^2)|\mathbf{E}_{\perp}+\hat{\mathbf{k}}\times{}c\mathbf{B}|$.  Here $\mathbf{E}_{\perp}$ is the component of the electric field perpendicular to the particle's motion (in the direction of $\hat{\mathbf{k}}$ for gamma-ray photons and $\hat{\mathbf{v}} = \bm{\beta}c/|\mathbf{v}|$ for electrons \& positrons).  The three-vector form for $\eta$ shows the origin of the geometrical factor $\theta$. In the case of an underdense plasma being struck by a single high-intensity laser pulse the electrons are rapidly accelerated to $\sim{}c$ in the direction of propagation of the laser, the $\mathbf{E}_{\perp}$ and $\bm{\beta}\times{}c\bm{B}$ terms in $\eta$ almost exactly cancel and so $\theta \ll 1$, $\eta$ is small and emission of photons and pairs consequently reduced.  In the case of an ultra-relativistic beam of electrons colliding with the laser pulse the terms add and $\theta\approx2$, dramatically increasing the level of emission.  In the case of counter-propagating laser pulses $\theta\approx1$ and the situation is also favorable for gamma-ray photon and pair production.  A similar configuration is found in laser solid interactions, with the counter-propagating beam provided by the reflected wave.  We will compare the effectiveness of the counter-propagating beam and laser-solid configurations in producing gamma-rays and pairs in section \ref{plasma_QED}.

\subsection{Emission Rates \& Monte-Carlo Model}
  
For a plane electromagnetic wave the (spin \& polarization averaged) rates of photon production by electrons and positrons of energy $\gamma{}m_ec^2$ \& pair production production by photons of energy $\hbar\omega_{\gamma}$ are, respectively: $\lambda_{\gamma}(\eta) = (\sqrt{3}\alpha_fc/\lambda_c)(\eta/\gamma)h(\eta)$ \& $\lambda_{\pm}(\chi) = (2\pi\alpha_fc/\lambda_c)(m_ec^2/\hbar\omega_{\gamma})\chi{}T_{\pm}(\chi)$ \cite{Erber_66}.  $\alpha_f$ is the fine-structure constant, $\lambda_c$ is the Compton wavelength.  $h(\eta) = \int_0^{\eta/2}d\chi{}F(\eta,\chi)/\chi$, $F(\eta,\chi)$ is the quantum-corrected synchrotron spectrum \cite{Erber_66}.  $T_{\pm}(\chi)\approx{}0.16K_{1/3}^2(2/3\chi)/\chi$.

The quantum (stochastic) nature of the emission \cite{Shen_72} can be modelled using a Monte-Carlo technique \cite{Sokolov_09,Duclous_11}.  The cumulative probability that a particle emits when passing thorough a plasma of optical depth $\tau$ is $P(\tau)=1-e^{-\tau}$; $\tau=\int_0^t\lambda[\eta(t')]dt'$ and $\lambda$ is the reaction rate.  To determine the optical depth each particle traverses before emitting $P$ is assigned a random value between 0 and 1 and the equation for $P$ above is then inverted to yield $\tau$.  For each particle the optical depth evolves according to the rate equations above until $\tau$ is reached and emission occurs.  The cumulative probability that an electron or positron with parameter $\eta$ emits a photon with $\chi$ (i.e. an energy $\hbar\omega_{\gamma}=2\gamma{}m_ec^2\chi/\eta$) is $P_{\chi}(\eta,\chi)=[1/h(\eta)]\int_0^{\chi}d\chi'F(\eta,\chi')/\chi$.   When a photon creates a pair it is annihilated and its energy shared between the generated electron and positron.  The cumulative probability that the positron takes fraction $f$ of the energy (parameterized by $\chi$) is $P_f(f,\chi)=\int_0^{f}df'p_f(f',\chi)$ \cite{Daugherty_83}.  $P_{\chi}$ \& $P_f$ are assigned to the emitted photon or positron at random in the range [0,1].  The corresponding values of $\chi$ or $f$ are determined by inverting the equations for $P_{\chi}$ or $P_f$.  After emitting a photon the emitting electron or positron recoils, its momentum changing by $\Delta{}\mathbf{p}=-(\hbar\omega_{\gamma}/c)\hat{\mathbf{p}}$, providing the quantum equivalent of the radiation reaction force \cite{Sokolov_10,DiPiazza_10}.  Note that in the limit where $\Delta\mathbf{p}\ll \mathbf{p}$ many photons are emitted for an appreciable change in the particles energy and therefore the whole synchrotron spectrum is sampled, this is identical to a classical treatment where the particle instantaneously emits the entire synchrotron spectrum, thus the Monte-Carlo algorithm agrees with a classical treatment of radiation reaction in the classical limit.

\subsection{QED-PIC}

The basis of the PIC technique \cite{Dawson_62} is the representation of the plasma as macroparticles, each representing many real particles such that the number of macroparticles is amenable to simulation.  Particle interactions are mediated by: (1) interpolating the charge and current densities resulting from the positions and velocities of the macroparticles onto a spatial grid; (2) solving Maxwell's equations for the $\mathbf{E}$ \& $\mathbf{B}$ fields; (3) interpolating these fields onto the particle's positions and pushing the particles using the Lorentz force law.  The inclusion of the QED processes is simplified by: the fact that the macroscopic fields may be treated classically and therefore step (2) remains unchanged; the macroscopic fields are quasi-static and therefore the QED interactions are point-like, occur instantaneously on the timescale of the PIC code and are consequently not resolved by the code.  Therefore we include the QED emission processes as a new step (0).  

During emission macrophotons and macropairs are created.  The pairs are treated in an equivalent way to the original electrons in the PIC code.  The photons are treated as massless, chargeless macroparticles which propagate ballistically.  The placement of the QED emission step at (0) ensures that the feedback defining QED-plasmas is simulated self-consistently.  Radiation reaction exerts a drag force, altering the velocity of the electrons and positrons and therefore the current in the plasma; pair production acts as a current source.  The inclusion of the QED processes before the PIC code solves Maxwell's equations means that this change in the current is included when the fields are updated.  The updated fields are then passed back to the QED routines and used to calculate emission during the next time-step.  

\section{$>$10PW Laser-Solid Interactions}
\label{sims_sect}

\begin{figure*}
\centering
\includegraphics[scale=0.8]{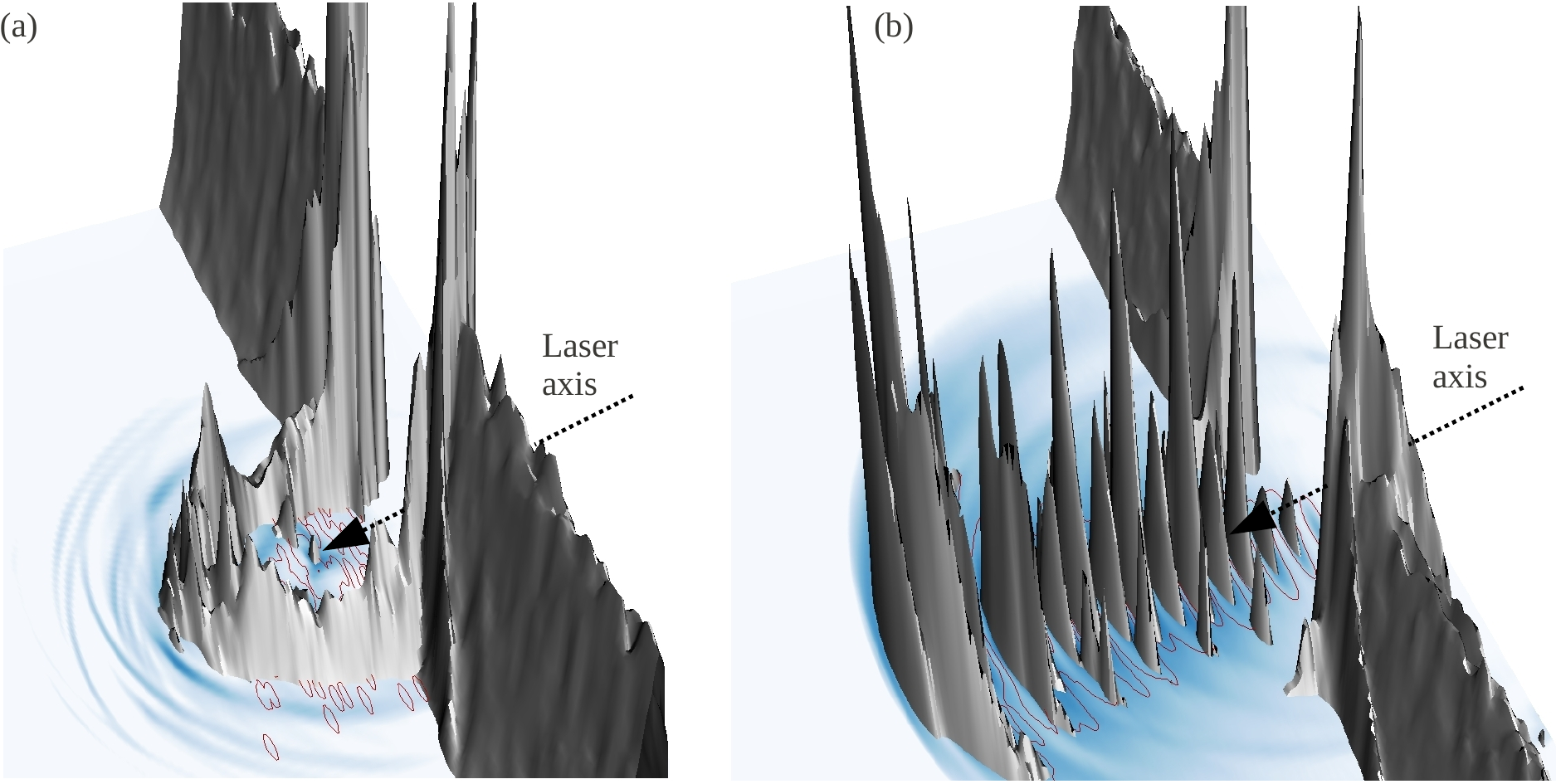}
\caption{\label{Sim_res} (Color online) (a) Gamma-ray photon (2D blue) \& positron (red contours) production in the interaction of a 12.5PW laser pulse with a solid aluminum target (3D grey).  (b) Equivalent plot for a 320PW laser pulse striking solid aluminum.}
\end{figure*}

Two-dimensional QED-PIC {\small{EPOCH}} simulations of 12.5PW and 320PW laser pulses striking solid aluminum targets at normal incidence have been performed which demonstrate the most important aspects of QED-plasma physics in both the radiation and QED dominated regimes.  In both cases the laser is linearly p-polarized and focussed to a spot of radius $1\mu$m on the target's surface [i.e. the spatial profile of the laser intensity here is $\propto\exp(-y^2/1\mu\mbox{m}^2)$].  Temporally the laser power $P=P_0$ for $0<t<30$fs and $P=0$ otherwise.  Therefore the intensity on-target is $4\times10^{23}$Wcm$^{-2}$ for $P_0=$12.5PW and $1\times10^{25}$Wcm$^{-2}$ for $P_0=$320PW.  The target is a fully-ionized aluminum foil of thickness $1\mu$m and initial density profile $\rho(x,y)=$2700kg$m^{-3}$ for $0<x<L$, $\rho(x,y)=0$ otherwise.  The target is discretized on a spatial grid with cell size 10nm and is represented by 1000 macroelectrons and 32 macroions per cell (12.5PW case) or 1857 macroelectrons and 142 macroions per cell (320PW case).

 Fig. \ref{Sim_res}(a) shows the results for the 12.5PW laser pulse.  This shows that prolific gamma-ray (2D blue) and positron (red contours) production occur as the laser bores into the solid (3D grey).  $4.8\times10^{13}$ gamma-ray photons with an average energy of 4.8MeV are produced, corresponding to 10\% conversion of laser energy to gamma-rays and therefore this interaction is in the radiation dominated regime.  $10^{10}$ positrons are produced.  Despite the large number generated the positrons are a minority species in the plasma, and therefore the sheath is generated by the `fast' electrons launched into the target \cite{Wilks_92}.  In this case the positrons pass through the sheath and readily escape the target.  A pure electron-positron plasma is formed behind the target with a maximum positron number density of $10^{26}$m$^{-3}$, $0.1$ times the non-relativistic critical density for optical lasers.  For comparison the highest positron density outside the target obtained in 1PW laser plasma experiments is $\sim10^{19}$m$^{-3}$.  It should be noted that similar numbers of pairs are generated in each case and that the dramatic increase in density is entirely due to the much smaller volume over which the pairs are generated in 10PW than in PW laser-plasma interactions ($\sim{}1\mu$m$^3$ compared to $\sim1$mm$^3$). 

The average positron energy is 320MeV.  This is much higher than the average photon energy and approximately twice that of the fast-electrons (140MeV).  This suggests that the positrons are born with relatively low energy and are rapidly accelerated by the laser to an energy equal to that of the fast-electrons and further accelerated by the sheath fields on leaving the target.  The sheath field acts to confine the fast-electrons inside the target and so accelerates positrons \cite{Chen_10}, doing work $\Phi{}m_ec^2$ approximately equal to the fast electron energy \cite{Ridgers_11}.  In this case we expect the average Lorentz factor of the positrons to be $\langle\gamma\rangle\approx a^{sol}_{HB}+\Phi\approx 2a^{sol}_{HB}=2eE_{HB}^{sol}\lambda_{lHB}/2\pi m_ec^2\approx$ 300MeV, which is consistent with the simulations.  In total 0.01\% of the laser energy is converted to positron energy and so their relative effect on the plasma dynamics is small.

Figure \ref{Sim_res}(b) shows simulation results for a 320PW laser striking a solid aluminum target.  $10^{16}$ gamma-ray photons and $10^{13}$ positrons are produced with average energies of 92MeV \& 2.2GeV respectively.  The maximum positron density is $1.8\times10^{30}$m$^{-3}$, an increase of four orders of magnitude for only a factor of 25 increase in laser intensity.  40\% of the energy is converted to gamma-rays and 10\% to electron-positron pairs.  Therefore at this extreme laser intensity both gamma-ray photon and pair production are crucial to the plasma dynamics and therefore the interaction is in the QED dominated regime.  

\subsection{The Effect of Plasma Physics Processes on the QED Rates}
\label{plasma_QED}

The key feature of a QED-plasma is the feedback between QED and classical plasma physics processes.  In this section we discuss the effect plasma physics processes have on the rates of the QED reactions in laser solid interactions.  This is best illustrated by comparing pair production in the interaction of a laser of intensity $I$ with a solid target to the alternative configuration consisting of two counter-propagating laser pulses of intensity $I/2$ interacting with a low-density gas \cite{Bell_08,Nerush_11} as in the laser-gas case complicated plasma effects are less important.  The laser-solid configuration has the clear advantages that the peak electric field is double that of laser-gas case due to reflection and that the pulse interacts with a dense plasma so that many pairs and photons may be produced even when the rates of reaction are low. 
  
\begin{figure}
\centering
\includegraphics[scale=0.9]{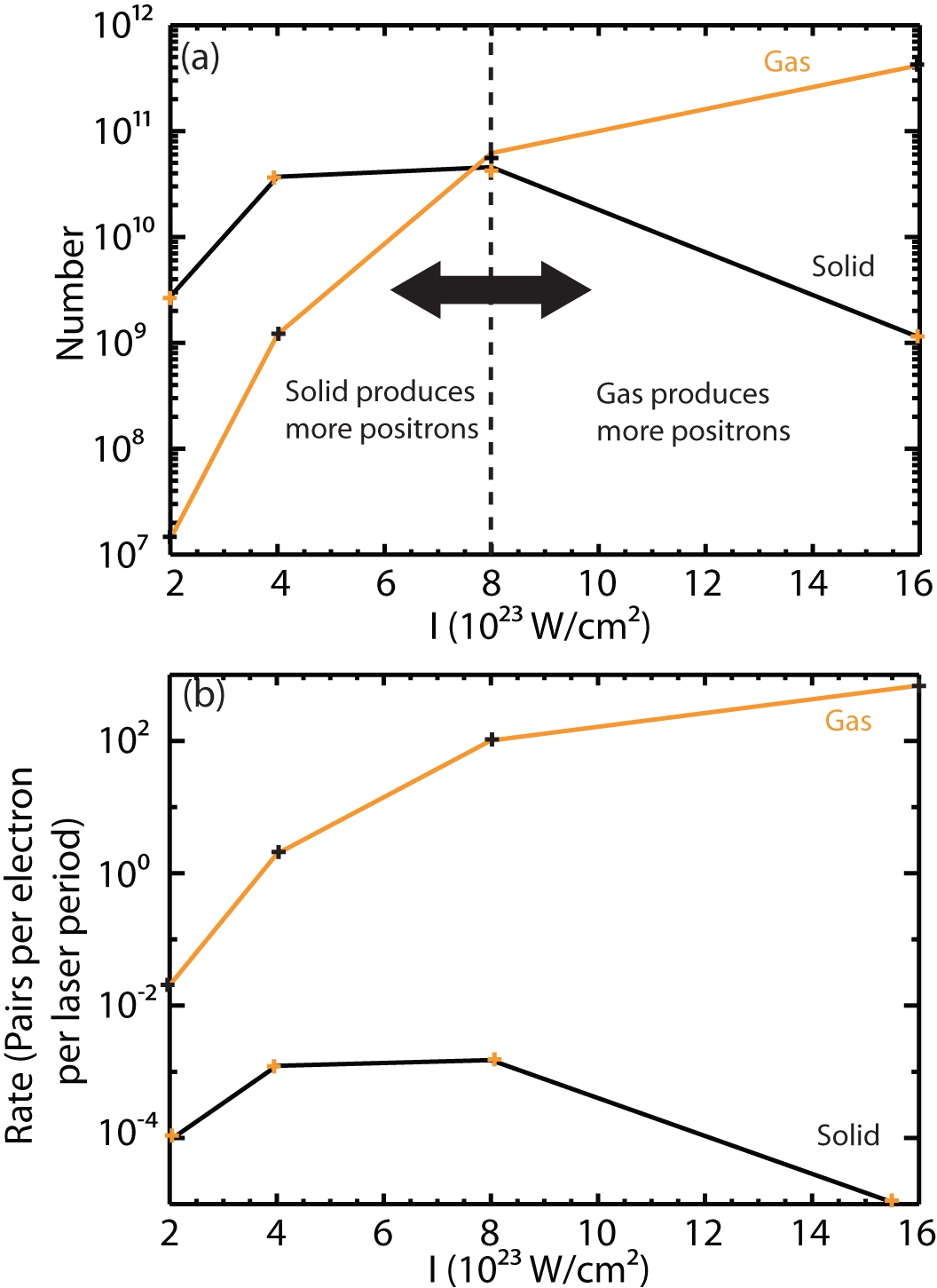}
\caption{\label{n_gamma_pairs} (Color online) (a) Number of positrons generated in the interaction of a laser pulse of intensity $I$ with solid and gas targets. (b) The rate of positron production in each of these cases.}
\end{figure}

A parameter scan of the effect of increasing laser intensity on the number of pairs produced by each configuration is conducted using one-dimensional {\small{\small{EPOCH}}} simulations.  The targets considered are: solid aluminum (density 2700kgm$^{-3}$) and a hydrogen gas-jet (density 0.02kgm$^{-3}$).  The solid targets are semi-infinite to avoid complications caused by the laser pulse breaking through the target.  Figure \ref{n_gamma_pairs}(a) shows the number of positrons produced by each configuration.  Due to the advantages previously mentioned the laser-solid case produces more positrons for $I<8\times10^{23}$Wcm$^{-2}$.  For $I>8\times10^{23}$Wcm$^{-2}$ the gas-jet configuration continues to behave as expected, with increasing intensity leading to increased pair production and when $\eta\sim1$ a large fraction of the pairs generated go on to produce additional pairs, the reaction runs away and a cascade of antimatter production ensues.  This is in good agreement with the results of Nerush \emph{et al} \cite{Nerush_11}.  In contrast pair production in the laser-solid case peaks when $I=8\times10^{23}$Wcm$^{-2}$ and then decreases for further increases in laser intensity.    
      
\begin{figure*}
\centering
\includegraphics[scale=0.85]{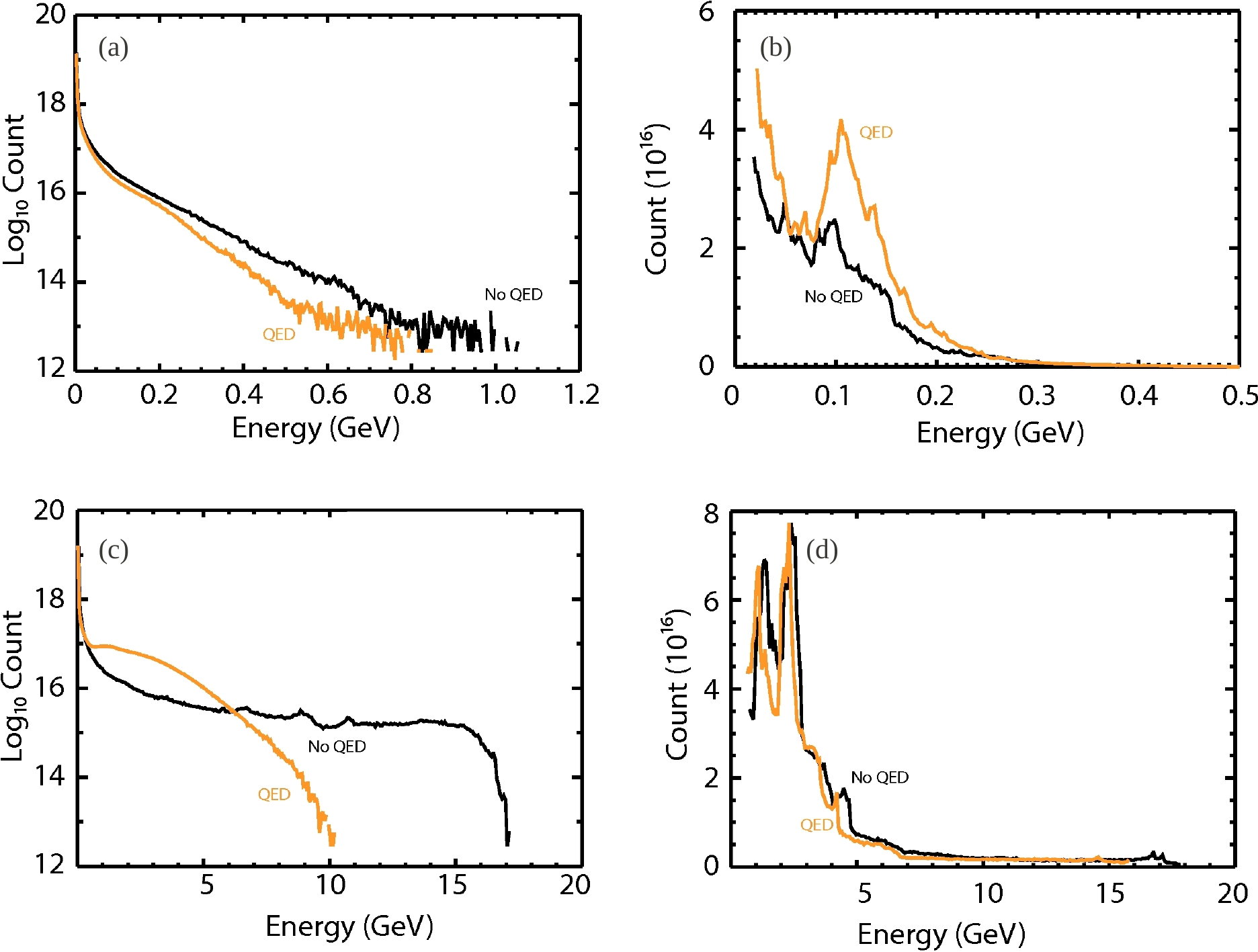}
\caption{\label{energy_specs} (Color online) Energy spectra of the plasma components with and without the inclusion of QED effects in both the 12.5PW and 320PW laser-solid simulations.  (a) \& (c) show electron spectra for 12.5PW and 320PW respectively, (b) \& (d) ion spectra for 12.5PW \& 320PW.}
\end{figure*}

The difference between the laser-solid and laser-gas configurations is more marked when considering the rate of pair production, shown in Figure \ref{n_gamma_pairs}(b).  The rate is substantially lower for the solid than the gas target at all intensities.  Several plasma effects have been proposed as being responsible for this reduction, namely: relativistic hole boring, the skin effect \& relativistic transparency \cite{Zhidkov_02,Ridgers_12}.  Qualitatively the reduction when $I<8\times10^{23}$Wcm$^{-2}$ is due to hole-boring \& the skin effect.  When the pulse strikes the solid surface its radiation pressure accelerates the surface to speed $v_{HB}=\beta_{HB}c$ in a process known as hole-boring.  The laser is reflected in the rest frame of the surface, in which its intensity is reduced by a factor of $(1-\beta_{HB})/(1+\beta_{HB})$, where $\beta_{HB}=\sqrt{\Xi}/(1+\sqrt{\Xi})$ and $\Xi=I/\rho{}c^3$ is the pistoning parameter for a laser of intensity $I$ and a target of density $\rho$ \cite{Robinson_09}.  From this formula one can see that in the 12.5PW interaction $v_{HB}\approx0.2c$ and the intensity in the rest frame of the surface is 0.7 times that in the lab frame.  Th electric field of the laser is evanescent inside the overdense solid and is reduced to $E^{sol}_{HB}=2(n_c/n_{eHB})^{1/2}E_{HB}^{max}$ (the skin effect).  $n_c=\gamma{}m_e\epsilon_0\omega_l^2/e^2$ is the relativistically corrected critical density for the plasma ($\gamma$ is the Lorentz factor to which the electrons are accelerated by the laser pulse); $E_{HB}^{max}$ is the peak laser electric field and $n_{eHB}$ the electron number density both in the hole-boring frame.  The reduction in the rate of pair production is consistent withe the reduction in the field in the solid target to $E^{sol}_{HB}$ \cite{Ridgers_12}. The Lorentz factor reached by the electrons in the solid is $\propto\sqrt{I}$, therefore if the laser intensity is increased eventually $n_c>n_e$ and the solid becomes underdense and therefore transparent \cite{Kaw_70}.  This occurs when $I>8\times10^{23}$Wcm$^{-2}$.  In this case the electrons are pushed forwards at $c$, the situation is similar to an underdense plasma illuminated by a single laser pulse and emission is drastically curtailed.  Therefore gamma-ray and pair production are maximized in laser-solid interactions when the solid is marginally overdense.  When this is the case the ratio  $n_e/n_c$ is minimized and this overcomes the effect of the increased reduction of $E_{HB}$ due to the increase in $v_{HB}$ (resulting from the decrease in $\rho$ and corresponding increase in $\Xi$).  This was shown with simple quantitative estimates in \cite{Ridgers_12}.  However some aspects of the model used there are inconsistent with the work in \cite{Chen_11} and further work is required.

The trend of decreasing levels of pair production with increasing intensity does not continue.  Prolific pair production was seen in the 320PW laser-solid interaction, despite the fact that at the intensity reached in this interaction ($10^{25}$Wcm$^{-2}$) the solid target is relativistically underdense.  This could be due to a new QED-mediated laser absorption mechanism which operates in underdense plasmas proposed by Brady \emph{et al} \cite{Brady_12}.  However care should be taken when comparing to the results in this paper, which discussed the radiation dominated regime.  In the QED dominated regime reached in the 320PW laser-solid interaction pair production is expected to lead to the generation of critical density pair plasmas over the duration of the laser pulse \cite{Nerush_11} which will clearly influence the plasma physics and as a result the macroscopic fields and therefore the QED rates.  For example, it has recently been shown that in this regime a dense pair-plasma can form in front of the solid surface \cite{Kirk_12}.  This pair plasma can absorb the laser.  In this case we expect the QED rates in the solid to be reduced.  Much more work is required to fully understand the influence of plasma physics processes on the emission rates in laser-solid interactions in the QED dominated regime.
   
%READ OVER PRL AND ENSURE WHAT I SAY HERE IS CONSISTEN

\subsection{The Effect of QED Emission on the Plasma Physics}

The substantial amount of energy converted to gamma-ray photons and pairs profoundly alters the laser energy absorbed by the electrons and ions in a QED-plasma and so the plasma processes which are driven by this energy.  First we consider the interaction of the 12.5PW laser pulse with the solid aluminum target.  In this case 10\% of the laser energy is converted to gamma-ray photons.  The effect that this has on the energy spectra of the electrons and ions is shown in Figures \ref{energy_specs}(a) \& \ref{energy_specs}(b).  The average energy of the fast electrons drops from 150MeV to 140MeV.  It is clear that the most energetic electrons are most affected as they emit photons most strongly.  Gamma-ray emission and the resulting radiation reaction causes a substantial difference in the ion spectrum, which develops two peaks.  Preliminary qualitative discussions of the modification of some plasma processes caused by this change in the energy spectra are given in Refs. \cite{Chen_11}.

Next consider the $I=1\times10^{25}$Wcm$^{-2}$ laser-solid interaction.  This interaction is in the QED dominated regime and a significant fraction of the laser energy is converted to both gamma-ray photons and pairs (40\% \& 10\% respectively).  Figure \ref{energy_specs}(c) shows the effect QED processes have on the electron energy.  The average fast electron energy is reduced from 4.5GeV to 2.0GeV.  As before, the high energy tail is preferentially damped.  However, in this case pair production is a significant source of electrons, which are generated at moderate energies and significantly enhance the spectrum here.  Figure \ref{energy_specs}(d) shows that QED effects do not significantly alter the ion spectrum.  In this underdense case the ions gain energy by coupling to electrostatic fields generated by the electrons as they are pushed forwards. as we have seen emission is not strong for electrons undergoing such motion.  We expect the dramatic modification of the electron and ion spectra caused by QED emission in the QED dominated regime to strongly effect the plasma processes; however, very little work has been done to elucidate this. 

\section{Conclusions}

Next generation high-power lasers, operating at intensities $>10^{23}$Wcm$^{-2}$, will generate a qualitatively new plasma state on interacting with matter.  These QED-plasmas are defined by feedback between QED emission processes and classical plasma physics effects.  We have described how the important QED processes: synchrotron-like gamma-ray photon emission \& multiphoton Breit-Wheeler pair production; can been included in a PIC code and used the resulting QED-PIC code to simulate this feedback in $>$10PW laser-solid interactions self-consistently.  We have shown that the rates of reaction in the simulations can only be explained when plasma effects are included and that the QED modifies these plasma effects by strongly altering the electron and ion energy spectra.  This alteration of the energy budget of laser-solid interactions may be important for proposed applications of 10PW lasers, such as ion acceleration or harmonic generation.  Simulation of a 12.5PW laser pulse striking a solid aluminum target demonstrates the conversion of a significant fraction (10\%) of the laser energy to gamma-ray photons.  In addition a pure electron-positron plasma is generated in the simulation with density seven orders of magnitude higher than currently achievable in laser-matter interactions.  Simulations of a 320PW laser-solid aluminum target interaction demonstrate that in this case we expect not only efficient (40\%) conversion of laser energy to gamma-rays, but also (10\%) to pairs.  This prolific production of gamma-ray photons and pairs may find application as an efficient and bright sources of these particles. 

\section*{Acknowledgements}

We acknowledge the support of the Centre for Scientific Computing, University of Warwick.  This work was funded by EPSRC grant numbers EP/GO55165/1 and EP/GO5495/1.


\begin{thebibliography}{99}
\bibitem{Mourou_07} G.A. Mourou, C.L. Labaune, M. Dunne, N. Naumova \& V.T. Tikhonchuk, Plasma Phys. Control. Fusion, \textbf{49}, B667 (2007)
\bibitem{Goldreich_69} P. Goldreich, \& W.H. Julian, ApJ., \textbf{157}, 869 (1969); R.D. Blandford, \& R.L. Znajek, MNRAS, \textbf{179}, 433 (1977)
\bibitem{Bula_96} C. Bula, K.T. McDonald, E.J. Prebys, C. Bamber, S.Boege, T. Kotseroglou, A.C. Melissinons, D.D. Meyerhofer, W. Ragg, D.L. Burke, R.C. Field, G. Horton-Smith, A,C, Odian, J.E. Spencer, D. Walz, S.C. Berridge, W.M. Bugg, K. Shmakov \& A.W. Weidermann, Phys. Rev. Lett., \textbf{76}, 3116 (1996); D.L. Burke, R.C. Field, G. Horton-Smith, J.E. Spencer, D. Walz, S.C. Berridge, W.M. Bugg, K.Shmakov, A.W. Weidemann, C. Bula, K.T. McDonald, E.J. Prebys, C. Bamber, S.J. Boege, T. Koffas, T. Kotseroglou, A.C. Melissinos, D.D. Meyerhoffer, D.A. Reis \& W. Ragg, Phys. Rev. Lett., \textbf{79}, 1626 (1997); U.I. Uggerh{\o}j, Rev. Mod. Phys., \textbf{77}, 1131 (2005)
\bibitem{Erber_66} T. Erber, Rev. Mod. Phys., \textbf{38}, 626 (1966); V.I. Ritus, J. Russ. Laser Res., \textbf{6}, 5 (1985)
\bibitem{Chen_10} H. Chen, S.C. Wilks, D.D. Meyerhofer, J. Bonlie, C.D. Chen, S.N. Chen, C. Courtois, L. Elberson, G. Gregori, W. Kruer, O. Landoas, J. Mithen, J. Myatt, C.D. Murphy, P. Nilson, D. Price, M. Schneider, R. Shepherd, C. Stoeckl, M. Tabak, R. Tommasini \& P. Beiersdorfer, Phys. Rev. Lett., \textbf{105}, 015003 (2010)
\bibitem{Bell_08} A.R. Bell, \& J.G. Kirk, Phys. Rev. Lett., \textbf{101}, 200403 (2008); A.M. Fedotov, N.B. Narozhny, G. Mourou, \& G. Korn, Phys. Rev. Lett., \textbf{105}, 080402 (2010); I.V. Sokolov, N.M. Naumova, J.A. Nees \& G. Mourou, Phys. Rev. Lett., \textbf{105}, 195005 (2010)
\bibitem{Ridgers_12} C.P. Ridgers, C.S. Brady, R. Duclous, J.G. Kirk, K. Bennett, T.D. Arber, A.P.L. Robinson \& A.R. Bell, Phys. Rev. Lett., \textbf{108}, 165006 (2012)
\bibitem{Sauter_31} F. Sauter, Z. Phys. \textbf{69}, 742 (1931); W. Heisenberg \& H. Euler, Z. Phys. \textbf{98}, 714 (1936); J. Schwinger, Phys. Rev., \textbf{82}, 664 (1951)
\bibitem{Sokolov_10} I.V. Sokolov, J.A. Nees, V.P. Yanovsky, N.M. Naumova \& G.A. Mourou, Phys. Rev. E, \textbf{81}, 036412 (2010)
\bibitem{Dirac_38} P.A.M. Dirac, Proc. R. Soc. A, \textbf{167}, 148 (1938); L.D. Landau \& E.M. Lifshitz, in The Course of Theoretical Physics Vol. 2 (Butterworth-Heinemann, Oxford, 1987), p222-229
\bibitem{Bahk_04} S.W. Bahk, P. Rousseau, T.A. Planchon, V. Chvykov, G. Kalintchenko, A. Maksimchuk, G.A. Mourou \& V. Yanovsky, Opt. Lett. \textbf{29}, 2837 (2004)
\bibitem{Shen_72} C.S. Shen \& D. White, Phys. Rev. Lett., \textbf{28}, 7 (1972)
\bibitem{Zhidkov_02} A. Zhidkov, J. Koga, A. Sasaki \& M. Uesaka, Phys. Rev. Lett., \textbf{88}, 185002 (2002); S. Kiselev, A. Pukhov \& I. Kostyukov, Phys. Rev. Lett \textbf{93}, 135004 (2003); N. Naumova, T. Schlegel, V.T. Tikhonchuk, C. Labaune, I.V. Sokolov \& G. Mourou, Eur. Phys. J. D, \textbf{55}, 393 (2009)
\bibitem{Chen_11}  M. Tamburini, F. Pegoraro, A. Di Piazza, C.H. Keitel \& A. Macchi, New J. Phys., \textbf{12}, 123005 (2010); M. Tamburini, F. Pegoraro, A. Di Piazza, C.H. Keitel, T.V. Liseykina \& A. Macchi, Nucl. Inst. \& Meth. in Phys. Res. A, \textbf{653}, 181 (2011); M. Chen, A. Pukhov, T. Yu \& Z. Sheng, Plasma Phys. Control. Fusion, \textbf{53}, 014004 (2011)
\bibitem{Nakamura_12} T. Nakamura, J.K. Koga, T. Zh. Esirkepov, M. Kando, G. Korn \& S.V. Bulanov, Phys. Rev. Lett., \textbf{108}, 195001 (2012)
\bibitem{Sokolov_09} I.V. Sokolov, N.M. Naumova, J.A. Nees, G.A. Mourou \& V.P. Yanovsky, Phys. Plasmas, \textbf{16}, 093115 (2009); I.V. Sokolov, N.M. Naumova \& J.A. Nees, Phys. Plasmas, \textbf{18}, 093109 (2011) 
\bibitem{Timhokin_10} A.N. Timokhin, MNRAS, \textbf{408}, 2092 (2010)
\bibitem{Nerush_11} E.N. Nerush, I.Y. Kostyukov, A.M. Fedotov, N.B. Narozhny, N.V. Elkina \& H. Ruhl, Phys. Rev. Lett., \textbf{106}, 035001 (2011)
\bibitem{Brady_11} C.S. Brady \& T.A. Arber, Plasma Phys. Control. Fusion, \textbf{53}, 015001 (2011)
\bibitem{Esirkepov_04} T. Esirkepov, M. Borghesi, S.V. Bulanov, G. Mourou \& T. Tajima, Phys. Rev. Lett., \textbf{92}, 175003 (2004)
\bibitem{Robinson_09} A.P.L. Robinson, P. Gibbon, M. Zepf, S. Kar, R.G. Evans \& C. Bellei, Plasma Phys. Control. Fusion, \textbf{51}, 024004 (2009)
\bibitem{Dromey_06} B. Dromey, M. Zepf, A. Gopal, K. Lancaster, M.S. Wei, K. Krushelnick, M. Tatarakis, N. Vakakis, S. Moustaizis, R. Kodama, M. Tampo, C. Stoeckl, R. Clarke, H. Habara, D. Neely, S. Karsch \& P. Norreys, Nature Phys., \textbf{2}, 456 (2006); T. Baeva, S. Gordienko \& A. Pukhov, Phys. Rev. E, \textbf{74}, 046404 (2006); A. Tarasevitch, K. Lobov, C. W\"{u}nsch \& D. von der Linde, Phys. Rev. Lett, \textbf{98}, 103902 (2007)
\bibitem{Heinzl_11} T. Heinzl, Int. J. Mod. Phys. Conf. Ser., \textbf{14}, 127 (2012).
\bibitem{Bagrov_90} V.G. Bagrov \& D.M. Gitman, in Exact Solutions of Relativistic Wave Equations (Kluwer Academic Publishers, Dordrecht, 1990) p43-114
\bibitem{Furry_51} W.H. Furry, Phys. Rev., \textbf{81}, 115 (1951)
\bibitem{Kirk_09} J.G. Kirk, A.R. Bell \& I. Arka, Plas. Phys. Control Fusion, \textbf{51}, 085008 (2009)
\bibitem{Duclous_11} R. Duclous, J.G. Kirk \& A.R. Bell, Plasma Phys. Control. Fusion, \textbf{53}, 015009 (2011); N.V. Elkina, A.M. Fedotov, I.Y. Kostyukov, M.V. Legkov, N.B. Narozhny, E.N. Nerush \& H. Ruhl, Phys. Rev. ST. AB., \textbf{14}, 054401 (2011)
\bibitem{Daugherty_83} J.K. Daugherty \& A.K. Harding, Ap. J., \textbf{273}, 761 (1983)
\bibitem{DiPiazza_10} A. DiPiazza, K.Z. Hatsagortsyan \& C.H. Keitel, Phys. Rev. Lett., \textbf{105}, 220403 (2010)
\bibitem{Dawson_62} J.M. Dawson, Phys. Fluids, \textbf{5}, 445 (1962); C.K. Birdsall \& A.B. Langdon, Plasma physics via computer simulation (McGraw-Hill, New York, 1985)
\bibitem{Wilks_92} S.C. Wilks, W.L. Kruer \& A.B. Langdon, Phys. Rev. Lett. \textbf{69}, 1383 (1992)
\bibitem{Ridgers_11} C.P. Ridgers, M. Sherlock, R.G. Evans, A.P.L. Robinson \& R.J. Kingham, Phys. Rev. E, \textbf{83}, 036404 (2011)
\bibitem{Kaw_70} P. Kaw \& J. Dawson, Phys. Fluids, \textbf{13}, 472 (1970)
\bibitem{Brady_12} C.S. Brady, C.P. Ridgers, T.D. Arber, A.R. Bell \& J.G.Kirk, Phys. Rev. Lett., \textbf{109}, 245006 (2012)
 \bibitem{Kirk_12} J.G. Kirk, A.R. Bell \& C.P. Ridgers, "Electron-positron pair modification of the hole-boring scenario in intense laser-solid interactions", submitted to \emph{Plasma Phys. Control. Fusion} 
\end{thebibliography}
\end{document}